\documentclass[aps,prl,twocolumn,showpacs,superscriptaddress,preprintnumbers,amsmath,amssymb]{revtex4-1}
\usepackage{graphicx}
\usepackage{color}
\draft 

\begin{document}


\title{Anomalous metallic state and anisotropic multiband superconductivity in Nb$_3$Pd$_{0.7}$Se$_7$} 



\author{Q. Zhang}
\affiliation{National High Magnetic Field Laboratory, Florida
State University, Tallahassee-FL 32310, USA}
\author{D. Rhodes}
\affiliation{National High Magnetic Field Laboratory, Florida
State University, Tallahassee-FL 32310, USA}
\author{B. Zeng}
\affiliation{National High Magnetic Field Laboratory, Florida
State University, Tallahassee-FL 32310, USA}
\author{T. Besara}
\affiliation{National High Magnetic Field Laboratory, Florida
State University, Tallahassee-FL 32310, USA}
\author{T. Siegrist}
\affiliation{National High Magnetic Field Laboratory, Florida
State University, Tallahassee-FL 32310, USA}
\affiliation{Department of Chemical and Biomedical Engineering, Florida State University, Tallahassee, Florida 32310, USA.}
\author{M. D. Johannes}
\affiliation{Center for Computational Materials Science, Naval Research Laboratory, Washington, DC 20375, USA}
\author{L. Balicas}
\email[]{balicas@magnet.fsu.edu}
\affiliation{National High Magnetic Field Laboratory, Florida
State University, Tallahassee-FL 32310, USA}

\date{\today}

\begin{abstract}
We report the discovery of superconductivity in Nb$_3$Pd$_{x}$Se$_7$ with a $x$-dependent superconducting transition-temperature as high as $T_c \simeq 2.1 $ K for $x \simeq0.7$ (middle point of the resistive transition).
Needle-like single crystals display anisotropic upper-critical fields with an anisotropy $\gamma = H^{b}_{c2}/H^{a}_{c2}$ as large as 6 between fields
applied along their needle axis (or $b-$axis) or along the $a-$axis. As for the Fe based superconductors $\gamma$ is temperature-dependent suggesting that Nb$_3$Pd$_{0.7}$Se$_7$ is a multi-band superconductor.
This is supported by band structure calculations which reveal a Fermi surface composed of quasi-one-dimensional and quasi-two-dimensional sheets of hole character, as well as three-dimensional sheets of
both hole- and electron-character. Remarkably, $H^{b}_{c2}$ is observed to saturate at $H^{b}_{c2}(T \rightarrow 0 \text{K}) \simeq 14.1$ T which is $4.26 \times H_p$ where $H_p$ is the Pauli-limiting
field in the weak-coupling regime. The synthesis procedure yields additional crystals belonging to the Nb$_2$Pd$_{x}$Se$_5$ phase which also becomes superconducting when the fraction of Pd is varied.
For both phases we find that superconductivity condenses out of an anomalous metallic state, i.e. displaying $\partial \rho/ \partial T < 0$ above $T_c$ similarly to what is observed in the
pseudogap-phase of the underdoped cuprates. An anomalous metallic state, low-dimensionality, multi-band character, extremely high and anisotropic $H_{c2}$s,
are all ingredients for unconventional superconductivity.

\end{abstract}


\maketitle 


\section{Introduction}

The Fe chalcogenide compounds, such as Fe$_{1+\delta}$Se, Fe$_{1+\delta}$Se$_x$Te$_{1-x}$, or the \emph{A}$_x$Fe$_{2-y}$Se$_2$ (\emph{A} = Tl, Rb or K) display remarkable superconducting properties.
For example, the tetragonal phase of Fe$_{1+\delta}$Se (i.e. when $0.01 \leq \delta \leq 0.04)$ displays superconductivity at $T_c \sim 8.5$ K only when $\delta = 0.01$  \cite{mcqueen},
although it was recently shown that intercalation with Li$_x$(NH$_2$)$_y$(NH$_3$)$_{1-y}$ or K leads to a dramatic enhancement in $T_c$ up to 43 K \cite{lucas} and 44 K \cite{zhang}, respectively.
For the Fe$_{1+\delta}$Se$_x$Te$_{1-x}$ series, recent angle resolved photoemission studies suggest a ratio for the superconducting gap to the Fermi energy $\Delta/ \varepsilon_F \approx 0.5$ placing
this system at the BCS to BEC crossover, or equivalently, that these are very strongly coupled superconductors \cite{kanigel}.
Finally, the \emph{A}$_x$Fe$_{2-y}$Se$_2$ is claimed to be close to an orbital-dependent Mott transition \cite{yi} indicating a certain resemblance with the cuprates,
although several studies support the microscopic coexistence of superconductivity with magnetism and concomitant insulating states \cite{may}.

Recently, we reported the discovery of yet a new transition-metal based anisotropic multi-band superconductor, i.e. Nb$_2$Pd$_{0.81}$S$_5$, which
displays extremely large upper critical fields \cite{alan}. In contrast to the Fe chalcogenides, this compound crystallizes in the lower symmetry
space-group $C2/m$ which, according to band structure calculations, leads to a complex Fermi surface (FS) composed of both two-dimensional (2D)
and quasi-one-dimensional (Q1D) sheets.  The Pd stoichiometry is predicted to play a major role since the calculations indicate that it displaces
the Fermi-level favoring nesting among the Q1D FS-sheets for particular wave-vectors, and this would lead to electronic instabilities and
possibly to itinerant magnetism \cite{alan}. Therefore, the Pd fraction is expected to play a role similar to that of F or Co and K in the
``1111" and ``122" Fe-arsenides respectively, \cite{chen,sefat,gfchen} where variations in their stoichiometry is observed to induce
superconductivity or increase the superconducting transition temperature in detriment of antiferromagnetism.

Given its composition and electronic anisotropy, one could expect Nb$_2$Pd$_{0.81}$S$_5$ to display physical similarities
with transition-metal dichalcogenides such as 2$H$-NbS$_2$ or 1$T$-TaS$_2$, or some of the transition metal trichalcogenides as NbSe$_3$. The former two compounds display
a charge-density wave (CDW) transition followed by superconductivity at lower temperatures \cite{tissen,ritschel}. While the latter displays two transitions towards CDW-phases upon
cooling \cite{monceau}. However, we did not detect any sharp anomaly that could indicate a transition akin to a Peierls instability in Nb$_2$Pd$_{0.81}$S$_5$, hence
we have no evidence for the coexistence of superconductivity with a density-wave like state which is claimed to enhance the superconducting upper-critical fields ($H_{c2}$) in the aforementioned compounds \cite{ausloos}.
Furthermore, and despite the similar values of $T_c$, Nb$_2$Pd$_{0.81}$S$_5$ displays remarkably larger $H_{c2}$s (by a factor of two or more depending on orientation)
when compared to either NbS$_2$ \cite{pfalzgraf} or NbSe$_2$ \cite{sanchez} and, in contrast to both compounds, it also displays a clear concave down curvature in $H_{c2}(T)$ for $H \|b$-axis.
These observations, indicate few similarities between Nb$_2$Pd$_{0.81}$S$_5$ and the dichalcogenide or trichalcogenide compounds.
In addition, as we will see below, the properties of the metallic state of new compounds emerging from Nb$_2$Pd$_{0.81}$S$_5$ by replacing S with Se are non-Fermi liquid like in the region
of temperatures just above $T_c$ in contrast to the aforementioned compounds. This behavior is very prominent in a very narrow range of values for the fraction $x$ of Pd, or when $0.67 \leq x \leq 0.71$, where one observes the
emergence of superconductivity from the non-superconducting compounds.
\begin{figure*}[htb]
\begin{center}
\includegraphics[width = 17 cm]{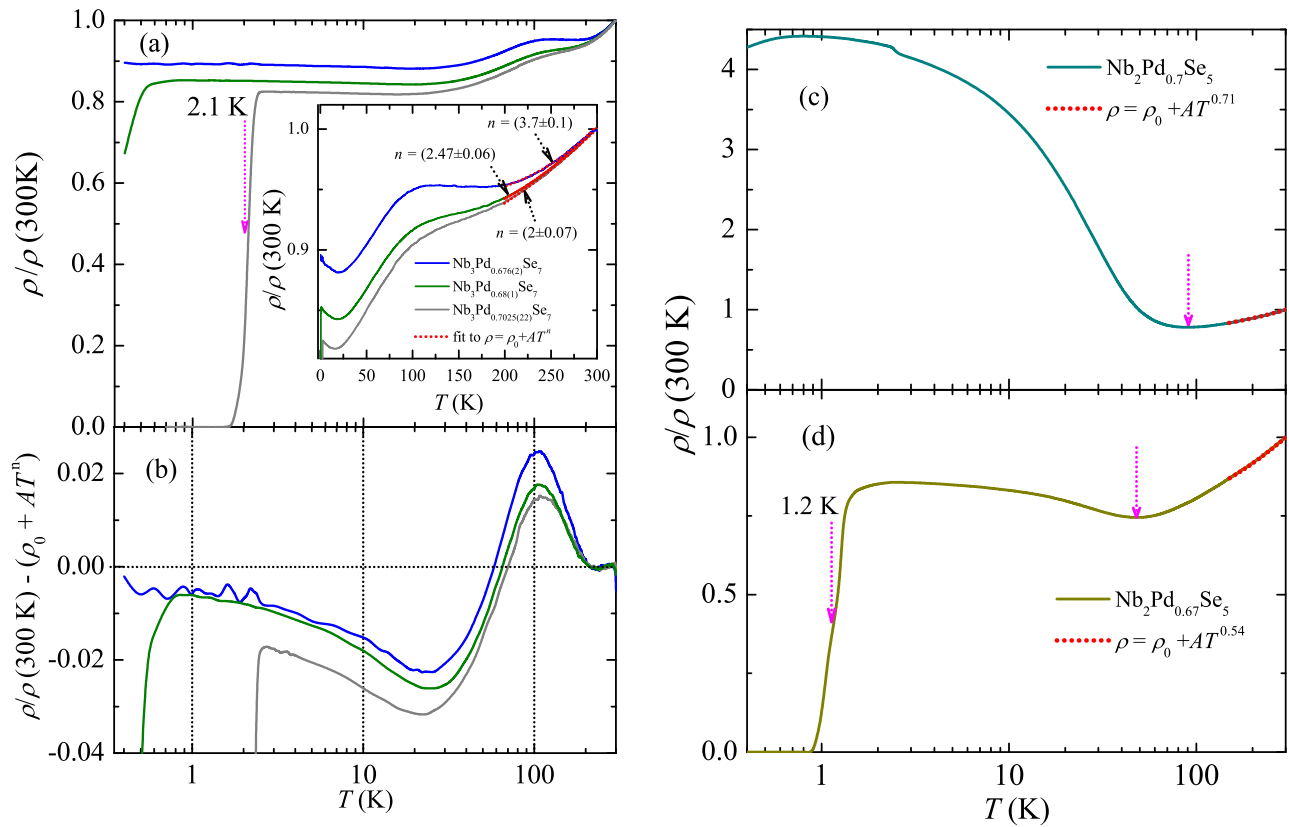}
\caption{(color online) (a) Resistivity $\rho$ normalized by its value at room temperature $\rho (300 \text{K})$ as a function of the temperature $T$ for three Nb$_3$Pd$_x$Se$_7$ single-crystals, i.e. $x = 0.676(2)$ (blue line),
0.68(1) (green line), and $\simeq 0.7$ (grey line), respectively. The dimensions of the superconducting crystal are $\sim 1\times10^{-4} \times 8.5\times10^{-3} \times 0.2$ cm$^{-3}$, leading to a room temperature resistivity of $\sim 70$ $\mu\Omega $cm (for a contacts separation of $\sim 0.05$ cm). Inset:  $\rho / \rho(300 \text{K}) $ as a function of $T$ in a linear temperature scale. Red lines are fits to the power law $\rho = \rho_0 + AT^n$ within the range $150 \leq T \leq 300$ K. The superconducting compound displays Fermi-liquid like behavior or $n=2$. (b) $\rho / \rho(300 \text{K}) $ after subtraction of the aforementioned power law.
Notice a sharp anomaly at $T^{\star} \sim 110$ K which is independent on the Pd content $x$, but whose amplitude is seen to decrease as $x$ increases.
This anomaly, of unknown origin is followed by a minimum in the resistivity at $T_{\text{min}} \sim 24$ K suggesting
electronic localization preceding the superconducting transition. (c) and (d) Resistivity $\rho / \rho (300 \text{K})$ as a function of the temperature $T$ for two Nb$_2$Pd$_x$Se$_5$
single-crystals, i.e. $x = 0.7$ and 0.67, respectively. Notice the observation of an upturn in the resistivity at a Pd-dependent temperature, indicating again an interplay with some form of electron
localization. The characteristic temperature $T_{\text{min}}$ below which $\rho$ becomes ``non-metallic", decreases as the Pd content decreases.
The dimensions of the Nb$_2$Pd$_{0.67}$Se$_5$ single-crystal are $\sim 1.75 \times10^{-4} \times 3.4\times10^{-3} \times 0.235$ cm$^{-3}$, leading to a room temperature resistivity of $\sim 225$ $\mu\Omega $cm.
Red lines are fits to the power-law behavior observed in the metallic state, which yield non-Fermi liquid like exponents $n$ ranging from $\sim 1$ (non-superconducting) to $\sim 0.4$ (in superconducting samples).}
\end{center}
\end{figure*}

In this manuscript, we show that new families of compounds can be created by varying the chemistry of Nb$_2$Pd$_{0.81}$S$_5$, leading to new superconducting compounds such
as Nb$_2$Pd$_x$Se$_5$ and Nb$_3$Pd$_x$Se$_7$. It turns out, as shown below, that Nb$_2$Pd$_x$Se$_5$ for $x \simeq 0.7$ displays a crossover from
metallic $(\partial \rho / \partial T >0)$ to ``non-metallic"  $(\partial \rho / \partial T < 0)$ upon cooling, therefore we show that superconductivity
in this compound emerges from an anomalous metallic state upon cooling.  This crossover like temperature is observed to decrease as $T_c$ increases (upon variation of $x$),
with the metallic state displaying non-Fermi liquid behavior as a function of the temperature, i.e. $\rho = \rho_0 +AT^n$ with $n$ ranging from $\sim 1$ to $\sim 0.4$ .
In contrast, the Nb$_3$Pd$_{0.7}$Se$_7$ compound, with a middle point superconducting transition temperature $T_c \simeq 2.1 $ K, do display Fermi-liquid like metallic behavior
(or $n=2)$ above 150 K. Nevertheless a mild anomaly centered around 110 K is observed in the resistivity, suggesting also a crossover or perhaps an electronic instability.
Although in this compound $T_c$ increases with \emph{x} or the Pd content, the position of this anomaly remains unaffected by it. Similarly to the Nb$_2$Pd$_x$Se$_5$ compounds, superconductivity
in the Nb$_3$Pd$_x$Se$_7$ series also emerges from a state displaying ``non-metallic" character  (or $\partial \rho / \partial T < 0$).

For these new chalcogenides, low dimensionality, multi-band effects, anomalous metallic behavior, and as will be discussed below for the case of
Nb$_3$Pd$_{0.7}$Se$_7$, anisotropic and extremely high upper critical fields, suggest an unconventional superconducting state. Here, we focus on
the superconducting phase-diagram of Nb$_3$Pd$_{0.7}$Se$_7$. A detailed account on the dependence of the
superconducting phase-diagrams of Nb$_2$Pd$_x$Se$_5$ and Nb$_3$Pd$_x$Se$_7$ on $x$ or the Pd content,
will be given elsewhere.

\section{Experimental}

Nb$_3$Pd$_{0.7}$Se$_7$ was grown via a solid state reaction under an Ar atmosphere: Nb (99.99 \%), Pd (99.99 \%),
and Se (99.999 \%) well mixed powders in the ratio of 4:1:10 were heated to a peak temperature of 850 $^{\circ}$C
at a rate of 100 $^{\circ}$C/h in sealed quartz ampoules, kept for 48 hours and subsequently quenched to room temperature.
The obtained single crystals formed thin long needles several millimeters in length, but with cross-sectional areas
ranging from $1 \times 5$ $\mu$m$^2$ up to $20 \times 100$  $\mu$m$^2$. The stoichiometric composition was determined by energy
dispersive \emph{X}-ray spectroscopy and single-crystal \emph{X}-ray structure refinement.
Two distinct crystallographic phases were identified as resulting from the growth conditions,
namely Nb$_3$Pd$_x$Se$_7$ and Nb$_2$Pd$_x$Se$_5$, with $0.67 \leq x \leq 0.701$ showing superconducting transition
temperatures $T_c$ ranging from $T_c < 0.3$ K up to $\sim 2.0$ and $\sim 1.0$ K, respectively.
The detailed results of our refinements, including the respective atomic positions for typical single-crystals of each phase is given in Appendix A.
A conventional four terminal configuration was used for the resistivity measurements which were performed under
magnetic-field either by using a Physical Parameter Measurements System (also for heat capacity measurements)
or a superconducting magnet coupled to a dilution refrigerator.
\begin{figure}[htb]
\begin{center}
\includegraphics[width = 8.6 cm]{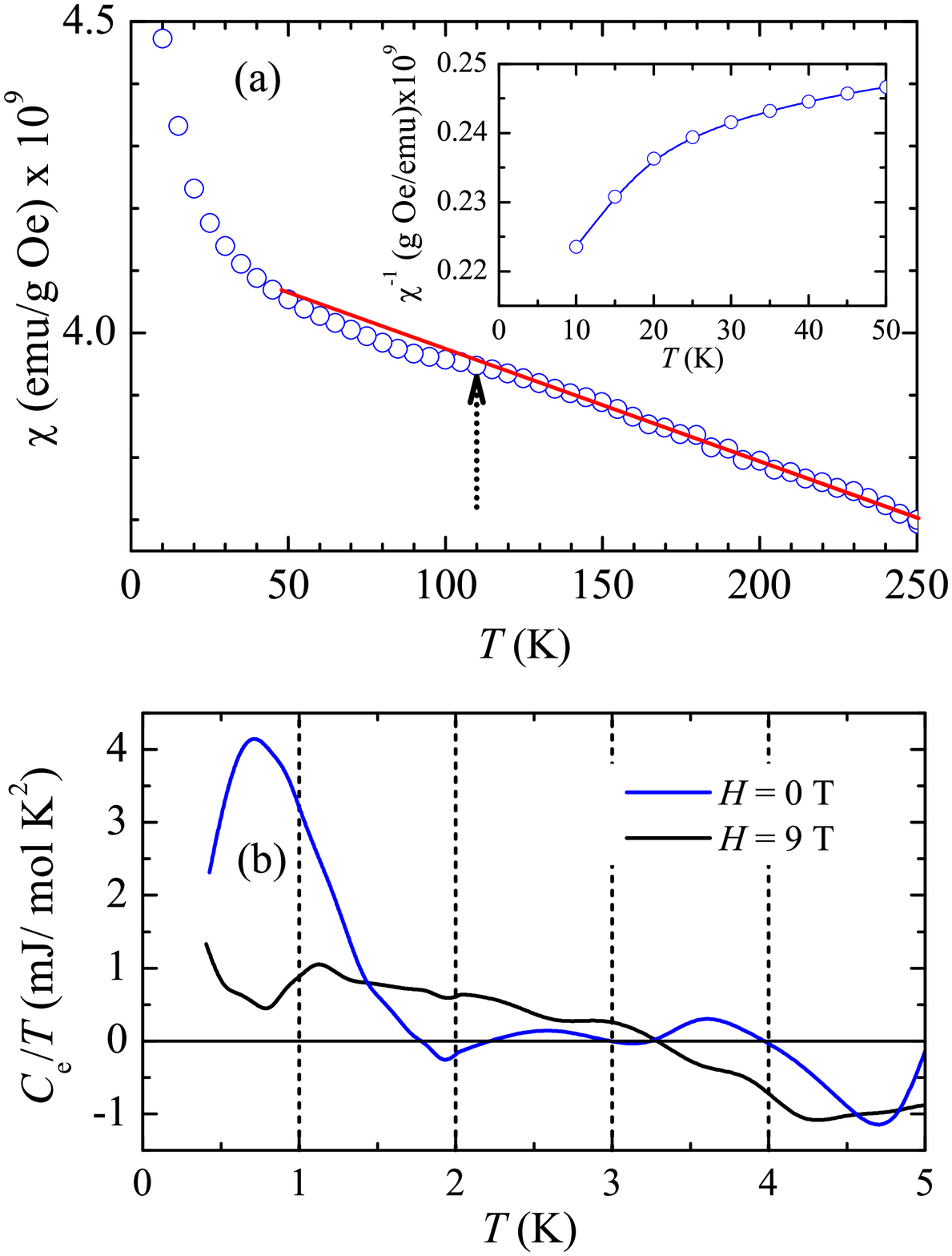}
\caption{(color online) Heat capacity normalized by $T$ and as a function of $T$ for hundreds of single-crystals from a typical synthesis batch, which contains both phases. One observes a broad anomaly centered around $T = 0.75$ K, which disappears with the application of a field of 9 T, indicating bulk superconductivity.}
\end{center}
\end{figure}
\begin{figure*}[htb]
\begin{center}
\includegraphics[width = 14 cm]{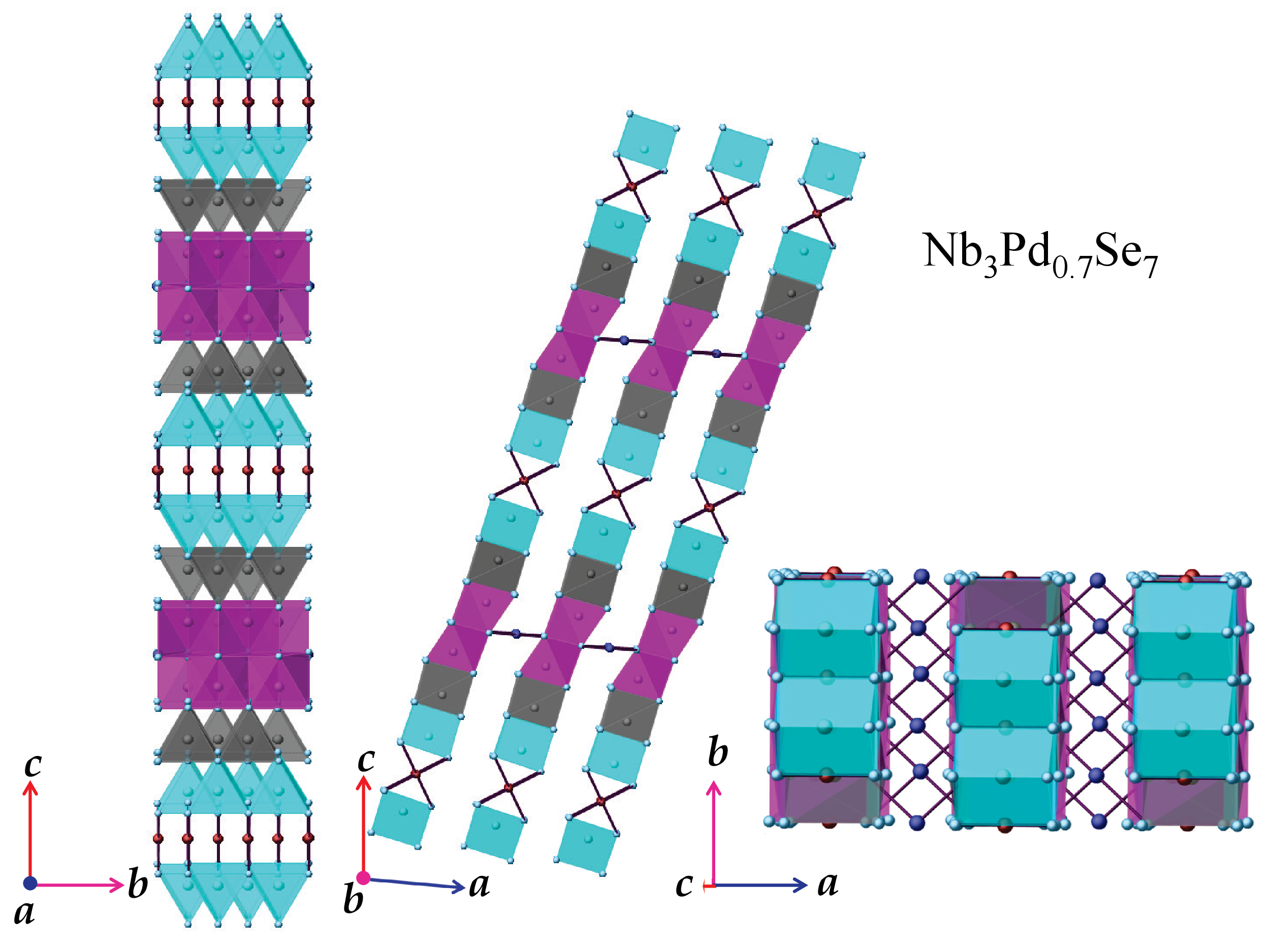}
\caption{(color online) (a) From left to right, distinct perspectives of the crystallographic structure of Nb$_3$Pd$_{0.7}$Se$_7$: along the $a-$, $b-$, and nearly along the $c-$axis, respectively. The Nb(1), Nb(2), and Nb(3) atoms are depicted in turquoise, grey, and magenta colors, respectively. Pd(1) and Pd(2) atoms are depicted in maroon and dark blue colors, whereas all the Se atoms are depicted in clear blue color, respectively.}
\end{center}
\end{figure*}

Figures 1 (a) to (d)  shows the resistivity $\rho$ normalized by its value at room temperature, i.e. $\rho/\rho(T = 300 \text{K})$ as a function of the temperature (in a logarithmic scale) for five single-crystals resulting from the same synthesis process. These crystals were analyzed through single-crystal $X$-ray diffraction measurements to reveal both their composition and their crystallographic structure. Traces in Figs. 1 (a) and (b) were obtained from crystals belonging to the Nb$_3$Pd$_x$Se$_7$ phase, whose Pd content $x$ was refined as $x = 0.676(2)$ (blue trace), $x = 0.68(1)$ (green trace), and $x = 0.7025(22)$ (grey trace), respectively. The room temperature resistivity of these crystals was found to oscillate between several tenths of $\mu\Omega$cm to a few hundreds of $\mu\Omega$cm.  This variability in the actual values of the resistivity can be attributed to the error bars in the determination of their geometrical factors, particularly their thickness which, for example, is approximately 1 $\mu$m for the superconducting single-crystal shown in Figs. 1 (a) and (b). As seen in the inset of Fig. 1 (a), a broad anomaly is observed in the resistivity around $T=100$ K, indicating either an electronic phase-transition akin to a spin-density wave state seen in the Fe-pnictide superconductors \cite{liu} or a crossover towards a new electronic regime, such as the pseudogap observed in the cuprates \cite{timusk}. Notice also that just above the superconducting transition $\rho$ displays an uncharacteristic $T$-dependence for a metal, i.e. with a negative $\partial \rho / \partial T$, indicating the existence of a prominent quasiparticle scattering mechanism. Red lines are fits to the power law $\rho = \rho_0 + AT^n$ which yields exponents $n$ ranging from 3.7 for the non-superconducting sample, to 2 for the superconducting one, thus suggesting that the superconducting sample displays Fermi-liquid behavior already at high temperatures. The higher exponents for the non-superconducting samples result from the pronounced upturn in the resistivity observed around 110 K. The overall behavior seen in Figs. 1 (a) and (b), i.e. a pronounced change in slope at higher temperatures followed by an uncharacteristic ``non-metallic" like behavior at lower temperatures, is remarkably similar to what is seen in the underdoped regime of, for example, La$_{2-x}$Sr$_x$CuO$_4$ \cite{takagi}, although the resistivity in those compounds follows a linear dependence on temperature at higher temperatures.

Figures 1 (c) and (d) show $\rho/\rho(T = 300 \text{K})$ for two Nb$_2$Pd$_{x}$Se$_5$ single crystals, i.e. respectively for $x=0.7$ and $x =0.67$ resulting from the same synthesis process. Notice the pronounced but progressive increase in resistivity below $\sim 100$ K for the $x=0.7$ sample, suggesting again a continuous or second-order like electronic phase-transition or perhaps a crossover towards a pseudogap-like regime. Therefore, for \emph{both} compounds $x = 0.7$ defines a threshold concentration separating superconducting from non-superconducting samples. Remarkably, while the increase in $x$ beyond 0.7 induces superconductivity in Nb$_3$Pd$_x$Se$_7$ it is observed to \emph{suppress} superconductivity in Nb$_2$Pd$_x$Se$_5$. This suggests that superconductivity is stabilized by small displacements of the Fermi level, or equivalently that in both compounds the Fermi level is located in close proximity to a van Hove singularity. For the Nb$_2$Pd$_x$Se$_5$ compounds, we found that the exponent in the power law describing $\rho$ at high temperatures (150 K $ \leq T \leq$ 300 K) evolves from a value close to 1 when $x \sim 0.7$,  to $ n \sim 0.5$ in the superconducting samples (or when $x \simeq 0.67 $) which corresponds to non-Fermi liquid behavior, and cannot be easily ascribed to scattering by phonons. Furthermore, the superconducting transition for Nb$_2$Pd$_{0.67}$Se$_5$ is preceded by an upturn in the resistivity starting at $T = 48$ K. Therefore, for both compounds and for $0.67 \leq x \lesssim 0.7$ superconductivity condenses out of an unconventional metallic state.

As discussed below as well as in Ref. \onlinecite{alan}, the geometry of the FS of these compounds is complex and contains quasi-one-dimensional sheets which are the necessary ingredient for a Peierls-like instability which could lead to itinerant magnetism. The broad anomaly seen around 110 K in Nb$_3$Pd$_{0.7}$S$_7$ single-crystals could correspond to evidence for such a transition and therefore in Fig. 2 (a) we show the magnetic susceptibility $\chi$ as a function of the temperature for a batch containing hundreds of randomly oriented single-crystals. We used hundreds of crystals, given that a typical individual single-crystal,  whose data are shown above, weights only $ \sim 1$ $\mu$g. We found that this particular batch is composed almost exclusively of crystals belonging to the Nb$_3$Pd$_x$Se$_7$ phase. As seen, at higher temperatures $\chi(T)$ varies little showing an unexpected linear dependence in temperature which could be an indication for Pauli-like susceptibility, but with a temperature-dependent density of states at the Fermi level. Around 110 K one observes just a mild deviation from linearity as indicated by the arrow, hence it does not represent a clear evidence for an electronic or magnetic phase-transition. For $T < 50$ K, $\chi(T)$ displays a sharp upturn which, as seen in the inset, does not correspond to a Curie tail: the inset plots the inverse of $\chi(T)$ in a limited temperature, i.e. for $T < 50 $ K, and as seen it is not linear in temperature. Consequently, this upturn cannot be attributed to the presence of impurities and most likely corresponds to evidence for magnetic correlations. A fit to a power law yields a very small exponent suggesting a $\log-T$ divergence.
Fig. 2 (b), on the other hand, shows the electronic contribution to the heat capacity $C_e$ normalized by the temperature $T$, after the subtraction of a $T^3$ term and for two values of the field, respectively $H=0$ and 9 T. Again, hundreds of single-crystals from one synthesis batch ($\sim 2$ mg), therefore of varying stoichiometry, were used for these measurements.  A broad anomaly in $C_e/T$ emerges below $\sim 1.5$ K and peaks at $\sim 0.7$ K, with its width determined by the distribution of $T_c$s, as seen in the upper panel. This anomaly is suppressed by the application of an external magnetic-field, as expected for bulk superconductivity. Therefore, our observations clearly indicate bulk superconductivity. We emphasize that the data shown in Fig. 1 corresponds to crystals displaying among the highest $T_c$s within the batches synthesized by us. In contrast, the heat capacity yields an average value for $T_c$.
\begin{figure}[htb]
\begin{center}
\includegraphics[width = 8.6 cm]{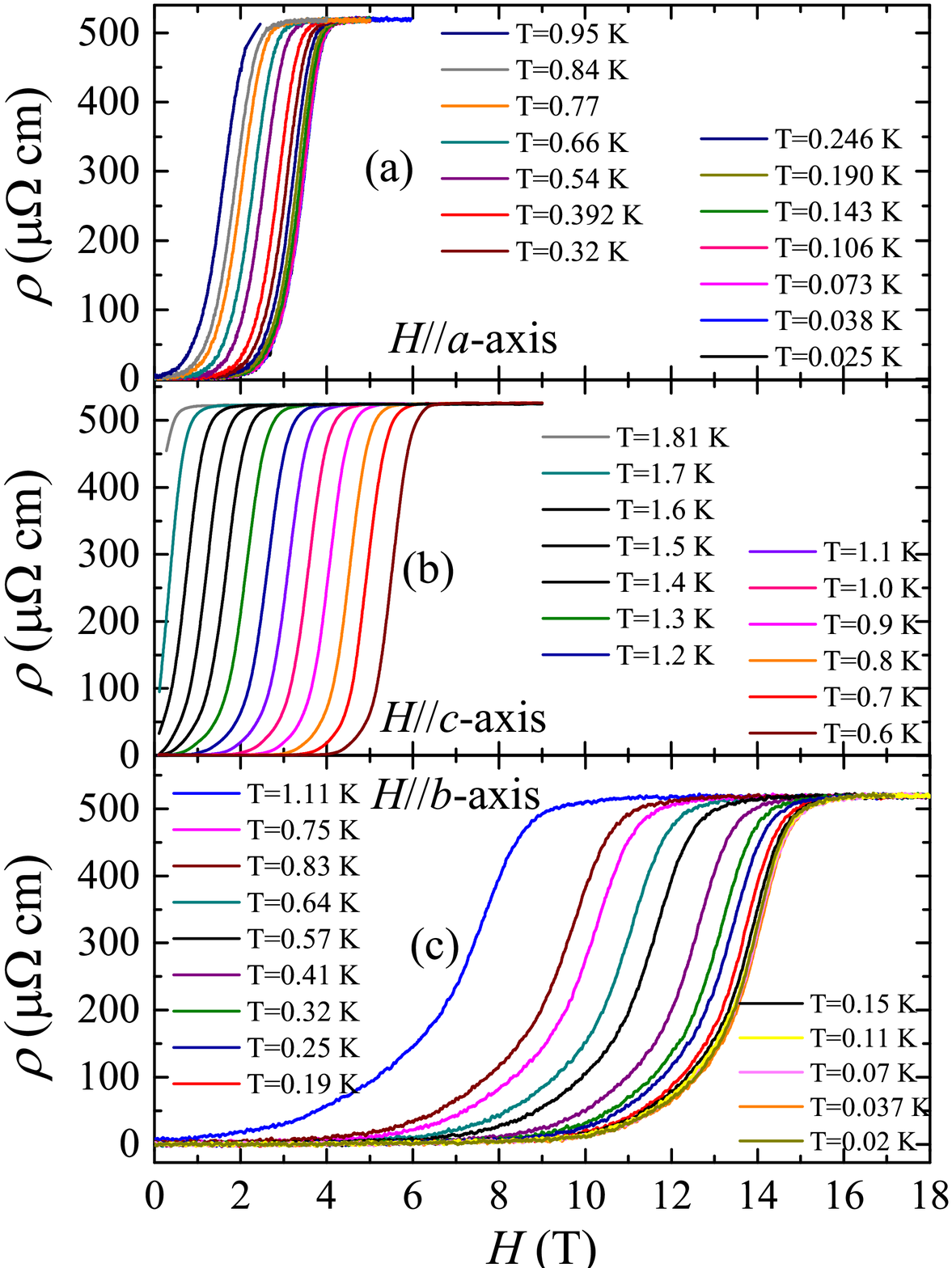}
\caption{(color online) (a) Resistivity $\rho$ as a function of the field $H$ applied along the $a-$axis of a Nb$_3$Pd$_{0.7}$Se$_7$ single-crystal for several temperatures. (b) Same as in (a) but fields along the $c-$ axis and at pumped $^3$He temperatures. (c) Same as in (a) but for fields along the $b-$axis.}
\end{center}
\end{figure}

Nb$_3$Pd$_{0.7}$Se$_7$ crystallizes in the space group $C2/m$ (see Fig. 3) and can be described as composed of sheets of a single Nb$_6$PdSe$_{14}$ basic unit which is comprised of chains formed by square-planar and trigonal-prismatic Se polyhedra which are approximately centered by Pd and Nb atoms, respectively (see Ref. \onlinecite{ibers}). Each chain extends along the [010] direction (needle axis). The Pd(1) atoms center a column of face-to-face square planes. Both Nb(1) and Nb(2) atoms occupy chains of edge-sharing trigonal prisms while the Nb(3) atoms occupy columns of face sharing trigonal prisms.
The basic Nb$_6$PdSe$_{14}$ is formed in such a way that the polyhedra in adjacent columns share their edges. The Nb$_6$PdSe$_{14}$ layers form via interdigitation of the Nb$_6$PdSe$_{14}$ units, with the Nb(3) atoms acquiring a seven-coordinate, monocapped trigonal-prismatic environment. The three-dimensional structure results from the stacking of these layers in a closed packed (Se atoms) fashion. The Pd(2) atom occupies a rhombic site between the layers and is square-planar coordinated with the Se(1) atoms.
\begin{figure}[htb]
\begin{center}
\includegraphics[width = 8.6 cm]{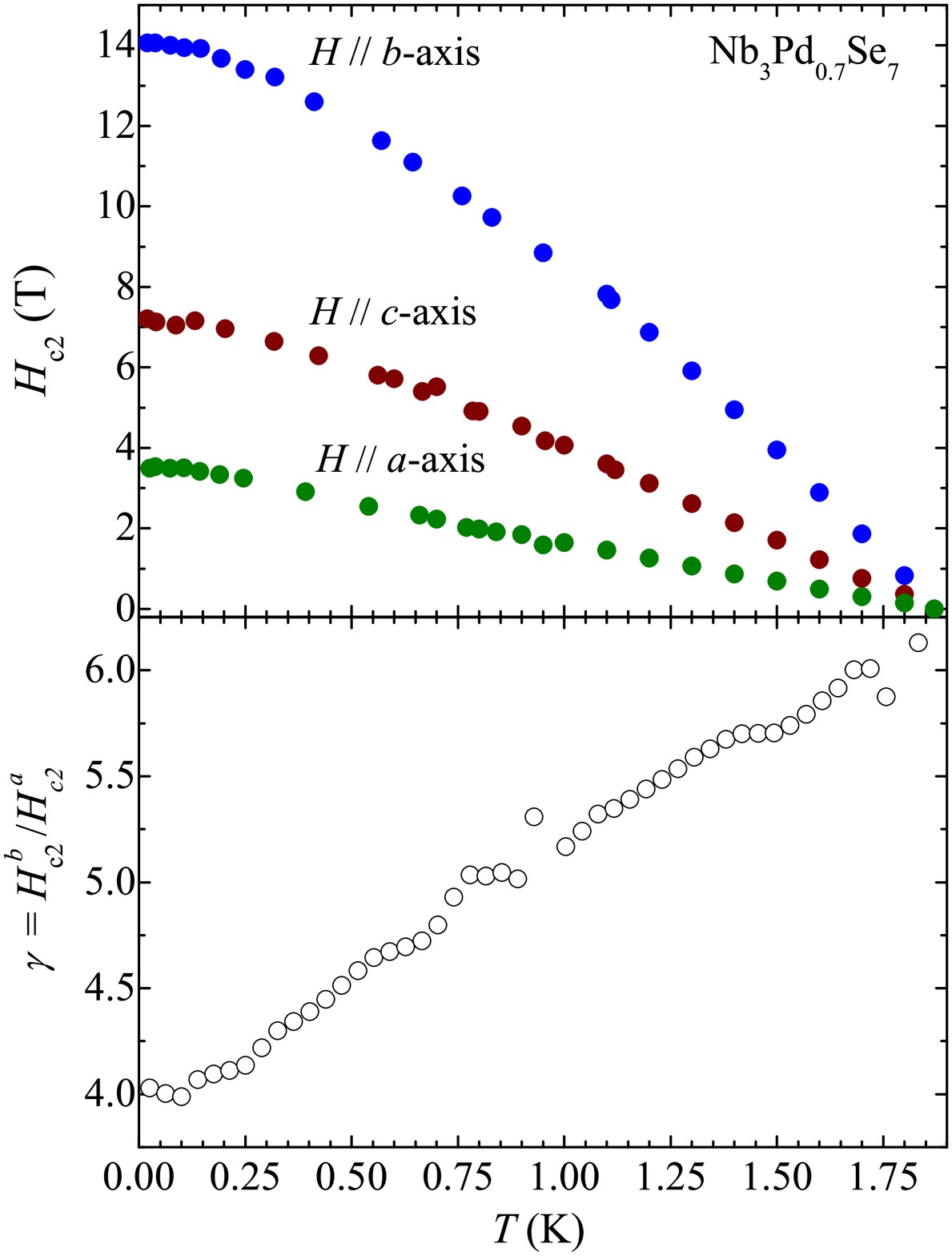}
\caption{(color online) Upper panel: phase boundary between superconducting and metallic states in the temperature $(T)$ - magnetic field $(H)$ plane for Nb$_3$Pd$_{0.7}$Se$_7$. Blue markers depict the phase-boundary for fields applied along the needle, or the $b-$axis of the crystal. Brown and green markers depict the phase-boundary for fields applied $c-$axis and $a-$axis, respectively. Lower panel: Anisotropy in upper-critical fields $\gamma = H^{b}_{c2}/H^{a}_{c2}$ as a function of temperature. Notice that $\gamma$ is temperature-dependent as previously observed in multi-band superconductors.}
\end{center}
\end{figure}
\begin{figure}[htb]
\begin{center}
\includegraphics[width = 8.6 cm]{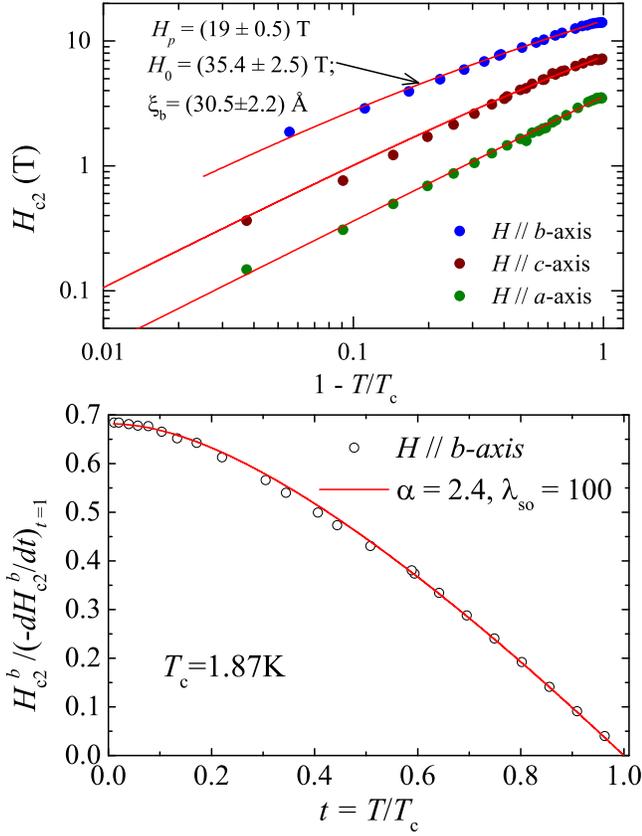}
\caption{(Color online) Upper panel: $H_{c2}$ for Nb$_3$Pd$_{0.7}$Se$_7$ as a function of $1 - t$, where $t = T/T_c$ for fields applied along the $b-$axis (blue dots), $c-$axis (brown dots) and $a-$axis (green dots).
Red lines are fits to Eq. 1.  Lower panel: $H_{c2}^b$ for fields along the $b-$axis normalized by the linear slope at $T_c$ and as a function of $t$. Red line is a fit to the WHH formalism. }
\end{center}
\end{figure}
\begin{figure}[htb]
\begin{center}
\includegraphics[width = 8.6 cm]{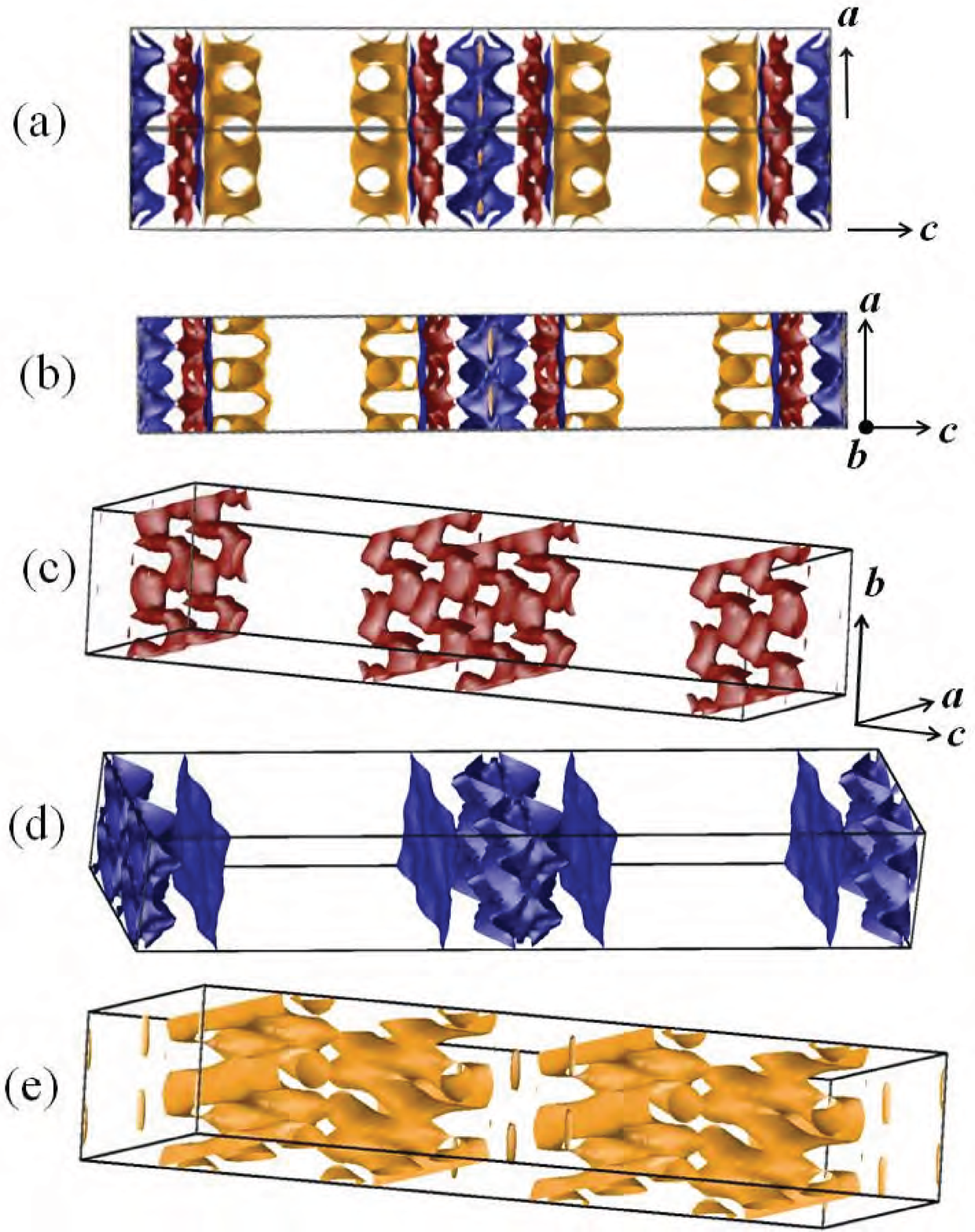}
\caption{(color online) (a) Top perspective of the geometry of the Fermi surface of Nb$_3$Pd$_{0.7}$Se$_7$ in a 2x2x2 conventional reciprocal unit
cell as obtained through DFT calculations. (b) Same as in (a) but through a lateral perspective. Maroon and blue FS sheets character, while the
gold intersecting cylinders are electron-like in one direction and hole-like in the other. Notice that the FS is composed of both
quasi-two-dimensional nearly cylindrical sheets as well as highly corrugated quasi-one-dimensional ones. The Q1D surfaces are extremely sensitive
to the Fermi energy, which is in turn sensitive to the exact Pd content.}
\end{center}
\end{figure}

Given the low crystallographic symmetry of Nb$_3$Pd$_{0.7}$Se$_7$ one expects an anisotropic response for magnetic fields applied along its distinct crystallographic axis.
Therefore, we show the resistivity of a second Nb$_3$Pd$_{0.7}$Se$_7$ single-crystal ($T_c \simeq 1.8$ K) for several temperatures and as a function of the magnetic-field $H$ applied along
the $a-$ and the $c-$axis (Figs. 4 (a) and 4 (b), respectively) and for fields along the $b-$ or the needle-axis (Fig. 4 (c)). This crystal was placed onto a sapphire substrate and the external field was initially applied perpendicularly to $(b,c)-$plane (Fig. 4 (a)). Subsequently, the field was rotated by 90$ ^{\circ}$ with respect to this previous orientation, while remaining perpendicular to needle-axis (Fig. 4 (b)). Subsequently, it was rotated by 90$^{\circ}$ around the $c-$axis in order to align the field along the $b-$axis (Fig. 4 (c)). As seen, the resistive transitions from the superconducting to the metallic state for the different field orientations reveal rather anisotropic upper-critical fields $H_{c2}$.

\section{Discussion}
In Figure 5, we display the resulting $H_{c2}$ - $T$ phase-diagram, constructed by plotting the middle point of the resistive transitions shown in Fig. 4 as well as from the resistivity as a function of $T$ under field (not shown). Remarkably, $H_{c2}$ for fields along the $b-$axis saturates at a value of $ H_{c2}(T \rightarrow 0 \text{ K}) \sim 14.1$ T which is $ 4.26 \times H_p [= 1.84 T_c (\simeq 1.8 \text{ K})]$, where $H_p$ is the Pauli limiting field in the weak coupling regime. To put this value in perspective, compare the ratio $H_{c2}(T \rightarrow 0 \text{ K})/T_c = 14.1 \text{ T}/1.8 \text{ K} = 7.83$ with the corresponding ratios for Fe$_{1+y}$Se$_{0.45}$Te$_{0.55}$, ($ \sim 50 \text{ T} / 14 \text{ K} \simeq 3.57$) \cite{tesfay}, CeCoIn$_5$ ($ \sim 12 \text{ T} / 2.3 \text{ K} = 5.2$) \cite{andrea}, URu$_2$Si$_2$  ($ \sim 12.5 \text{ T} / 1.5 \text{ K} = 8.33$) \cite{brison} which, according to all evidence, are unconventional and strongly-correlated superconductors. Notice that this ratio for Nb$_3$Pd$_{0.7}$Se$_5$ also surpasses the respective one for Nb$_2$Pd$_{0.81}$S$_5$ ($\sim$ 37 T/6.6 K $\simeq$ 5.6) \cite{alan}, or for the Chevrel-phase PbMo$_6$S$_8$ ($\sim$ 60 T/13.3 K $\simeq$ 4.51) \cite{okuda}, and obviously the ratio for the widely used Nb$_3$Sn compound ($\sim$ 30 T/18 K $\simeq$ 1.67) \cite{nb3sn}. Finally, and although
the anisotropy $\gamma$ for Nb$_3$Pd$_{0.7}$Se$_7$ is not as high as the one reported for Li$_{0.9}$Mo$_6$O$_{17}$, which suggests that this later compound is considerably more quasi-one-dimensional \cite{hussey}, its ratio $H_{c2}(T \rightarrow 0 \text{ K})/T_c \sim 15 \text{ T}/2.2 \text{ K} = 6.82$ still is smaller than the ratio reported here.

Below we show Ginzburg-Landau fittings of the phase-boundary, which for fields along the $b-$axis yields an orbital limiting field $H_0 = (35 \pm 3) $ T and $H_p = (19 \pm 1)$ T,
therefore suggesting that Nb$_3$Pd$_{0.7}$Se$_7$ could be a Pauli limited superconductor displaying a very large Maki parameter $\alpha =\sqrt{2}H_0/H_p = 2.6$ if one uses both field values resulting from the fittings,
or $\alpha \simeq 3.5$ if one uses the actual experimental result $14.1$ T. In any case, these values for $\alpha$ are $ > 1.8$ the value required for the observation of the Fulde-Ferrel-Larkin-Ovchinikov-state \cite{gruenberg}.
Below we also show a fit of $H_{c2}^b (T)$ to the Werthamer-Helfand-Hohenberg formalism \cite{whh}, indicating that a large $\alpha$ value of  $\sim 2.4$ would fit the boundary, but would require a large spin-orbit coupling implying that this coupling is relevant for this system. Finally, as seen in the lower panel of Fig. 4, the superconducting anisotropy $\gamma = H_{c2}^b/ H_{c2}^{a}$
is $T$-dependent as seen in the Fe pnictides/chalcogenides and interpreted as evidence for multi-band effects \cite{yuan, yamamoto}.

In order to evaluate the contributions of both orbital and Pauli pair-breaking effects for all field orientations
we analyze our $H_{c2} (T)$ data at temperatures close to $T_c$ where the Ginzburg-Landau (GL) theory yields \cite{alex}:
\begin{equation}
\left( \frac{H}{H_p}\right)^2 + \frac{H}{H_\text{o}} = 1 - \frac{T}{T_c}
\label{gl}
\end{equation}
Very close to the critical temperature, $(T_c - T)/T_c \ll (H_p/H_o)^2$, the first paramagnetic term in the left hand side is negligible and Eq. (1) yields the orbital linear Ginzburg-Landau temperature dependence,
$H_{c2} = H_o(1 - T/T_c)$.  At lower temperatures, $(T_c - T)/T_c > (H_p /H_o)^2$, the Pauli limiting field $H_p$ dominates the shape of $H_{c2} (T) \propto (1 - t)^{1/2}$ even in the GL domain if $H_p < H_o$.
The latter inequality is equivalent to the condition that the Maki parameter $\alpha \sim H_o/H_p > 1$ is large enough, assuring that the paramagnetic effects are essential.
Shown in the upper panel of Fig. 6 are the log-log plots of our $H_{c2}(T)$ as a function of $1 - T/T_c$ where the red lines are fits to Eq. (1).
As seen, for fields along the $b-$axis (blue markers) the fitting yields $H_p = (19 \pm 0.5)$ T and  $H_o = (35 \pm 3)$ T respectively, implying a Maki parameter of 2.6 if one uses both values resulting from the fittings,
or $\alpha \simeq 3.5$ if one used the experimental result $14.1$ T as a tentative value for the Pauli limiting field. In any case, such large values for the paramagnetic limiting field relative to $T_c$ would be difficult to understand for a conventional superconductor, suggesting either very strong correlations renormalizing $H_p$ as seen in the heavy-fermions, or an unconventional pairing mechanism, or both.
Notice that these values for $\alpha$ are $ > 1.8$ the value required for the observation of the FFLO-state \cite{gruenberg}.
For the other two orientations, one obtains similar values for both $H_p$ and $H_o$.
The lower panel of Fig. 6 shows a fit of $H_{c2}^b$ normalized by the slope of its linear dependence at $T_c$ and as a function of the reduced temperature $T/T_c$.
Red line is a fit to the Werthamer, Helfand, and Hohenberg formalism \cite{whh} with a tentative value of the Maki parameter $ \alpha = 2.4 $ which yields a extremely large value for the spin-orbit coupling
parameter $\lambda_{so} = 100$. Although it remains unclear how reliable such a large value is, this suggests that the spin-orbit coupling plays a relevant role for this system.

To shed some light about the possible superconducting pairing mechanism in Nb$_3$Pd$_{0.7}$Se$_5$ we have performed band structure calculations
to determine the geometry of the Fermi surface, the results are summarized in Fig. 7.
Density functional theory calculations using Wien2K \cite{wien2k} with the Generalized Gradient Approximation (GGA) \cite{gga} to the exchange correlation potential were employed to calculate the self-consistent energy eigenvalues at 16,000 k-points in the reciprocal lattice. A doubled cell with one Pd(2) atom removed was used in order to achieve a formula unit of Nb$_6$Pd$_{1.5}$Se$_{14}$, resulting in band-folding in the smaller reciprocal space cell.
The FSs were calculated using the experimental lattice constants and atomic positions established in this work.  The centering symmetry of space group \# 12 ($C2/m$) was eliminated so that half the Pd at the $2a$ Wyckoff position could be removed, resulting in a doubled unit cell with formula Nb$_6$Pd$_{1.5}$Se$_{14}$, {\it i.e.} with slightly more Pd than is incorporated experimentally.  As seen in Figs. 7 (c) and (d) the
resulting FS is composed of quasi-2D sheets (maroon surfaces) of hole character, and a
set of strongly and weakly warped Q1D (blue) sheets of hole character whose
geometry is sensitive to the exact Pd content. The complex 3D network
itself (golden) seen in Fig. 7 (e), contains both hole-and electron-like orbits. Therefore,
given that the resulting FS is composed of multiple sheets, having both electron- and
hole-character, this system can indeed be classified as a multi-band superconductor. Spin-polarized
calculations show that a ferromagnetic state is stable, with a small moment attributable to some
Nb atoms (~0.27 $\mu_B$/Nb(1), ~0.20 $\mu_B$/Nb(2)) and a polarized, delocalized interstitial
density.  The stabilization energy is extremely small, and manipulable via variations in the Pd
content indicating that the system is in proximity to magnetism. If a spin
fluctuation mechanism were in play, $T_c$ might be tunable by varying the stoichiometry as seen here.

\section{Summary}

In summary, Nb$_3$Pd$_{0.7}$Se$_7$ is a new chalcogenide based multi-band superconductor, displaying extremely large upper critical-fields, i.e. comparable to those of unconventional strongly correlated heavy-fermion superconductors. The synthesis process also yields crystals of the Nb$_2$Pd$_x$Se$_5$ phase which also displays superconductivity. Both compounds display an anomalous metallic behavior, particularly for temperatures above
their respective superconducting transitions which bears resemblances with the pseudogap phase of the cuprates. Such anomalous metallic behavior was not observed in Nb$_5$Pd$_{0.81}$S$_5$ which displays a considerably higher superconducting transition temperature \cite{alan}. This clearly indicates that the fraction $x$ of Pd plays the role of a dopant, i.e. by varying its content one can, for instance, suppress non-metallic states and stabilize superconductivity as observed in the Nb$_2$Pd$_x$Se$_5$ series. Notice that this study was confined to a narrow range in $x$, from $\sim 0.67$ to $\gtrsim 0.7$ where we have shown that small variations in $x$ or $\Delta x = 0.03$ easily suppress superconductivity in both families of compounds. Therefore, the overall behavior reported here for the Nb$_3$Pd$_x$Se$_7$ series is clearly characteristic of an underdoped regime. As we already implied within the Introduction, an anomalous metallic behavior with broad anomalies which cannot be clearly attributed to phase-transitions and which instead suggest crossovers towards new electronic regimes, makes these compounds particularly distinct from the dichalcogenides or the trichalcogenides. Notice, that in quasi-one-dimensional or quasi-two-dimensional systems such as these, both charge-density waves \cite{gruner1} coupling to lattice distortions, and spin-density waves \cite{gruner2} resulting from electronic correlations lead to sharp, first-order like anomalies in their physical properties such as the resistivity \cite{neven}, in contrast to what is seen here. In fact, it would seem that these compounds are more akin to the Fe chalcogenide superconductors: for example in the Fe$_{1+y}$Se$_x$Te$_{1-x}$ series the resistivity in the metallic state is known to display a $-\log T$ dependence above the superconducting transition which can be suppressed upon careful annealing \cite{tesfay}.

A complex Fermi surface composed also of quasi-one-dimensional sheets, interplay of superconductivity with an anomalous metallic state, extremely high upper critical fields coupled to multi-band behavior, suggest an unconventional pairing symmetry.

\section{Acknowledgement}
L.~B. is supported by DOE-BES through award DE-SC0002613. T. B. and T. S. are supported by DOE-BES through award DE-SC0008832, and by FSU.
Funding for M.D.J. was provided by the Office of Naval Research (ONR) through the Naval Research Laboratory's Basic Research Program.
The NHMFL is supported by NSF through NSF-DMR-0084173 and the State of Florida.

\appendix
\section{Appendix: Parameters of Nb$_2$Pd$_{0.67}$Se$_5$ and Nb$_3$Pd$_{0.7}$Se$_7$ as derived from the \emph{X}-ray refinements}

\begin {table}[htp]
\begin{center}
\begin{tabular}{|c|c|c|}
  \hline
  \textbf{Compound} & Nb$_2$Pd$_{0.67}$Se$_5$ & Nb$_3$Pd$_{0.70}$Se$_7$ \\
  \hline
  Space group & \emph{C2/m} & \emph{C2/m} \\
  \emph{a} (\AA) & 12.8325(7) & 12.7965(6) \\
  \emph{b} (\AA) & 3.39327(18) & 3.40591(17) \\
  \emph{c} (\AA) & 15.3859(8) & 21.0370(13) \\
  $\beta$ $(^{\circ}$) & 101.471(5) & 95.530(5) \\
  $\alpha$, $\gamma$ $(^{\circ}$) & 90.0 & 90.0 \\
  $Z$ & 4 & 4 \\
  $V$ (\AA$^3$)& 656.59(6) & 912.60(8) \\
  $d_{\text{cal.}}$ (g/cm$^3$) & 6.599 & 6.590 \\
  \emph{T} (K) & 298 & 298 \\
  Range $(^{\circ})$ & $3.24 < \theta$ & $2.92< \theta$ \\
  & $ <66.41$ & $<45.00$ \\
  Reflections & 6162 & 4170 \\
  Parameters refined & 52 & 70 \\
  Goodness-of-fit & 1.0046 & 1.1150 \\
  \hline
\end{tabular}
\caption{Crystallographic parameters as extracted from single-crystal \emph{x}-ray refinements
for two typical crystals resulting from the synthesis process.}
\end{center}
\end{table}

\begin {table}[htp]
\begin{center}
\begin{tabular}{|c|}
\hline
  Nb$_2$Pd$_{0.67}$Se$_5$  \\
\hline
\end{tabular}
\vspace{0.5 cm}
\begin{tabular}{|c|c|c|c|c|c|c|}
\hline
\hline
Atom &  Site & Occupation & $x$ & $y$ & $z$ & $U_{\text{eq}}$\\
\hline
Nb1& 4i	& 1&	0.07595(4)&	1/2&	0.17999(3)&	0.0099\\
Nb2 &4i&1&	0.15288(4)	&0	&0.37854(3)	&0.0078\\
Pd1	& 2a&1&	0	&0	&0	&0.0124\\
Pd2	& 2c&0.348(5)&	0	&0	&1/2	&0.0117\\
Se1	& 4i&1	&0.35036(5)	&0	&0.48914(3)	&0.0098\\
Se2	& 4i&1	&0.25374(4)	&1/2&	0.29566(3)	&0.0086\\
Se3	& 4i&1	&0.17536(5)	&0	&0.09778(4)	&0.0106\\
Se4	& 4i&1	&0.42303(5)	&1/2&	0.13214(4)	&0.0104\\
Se5	& 4i&1	&0.50041(4)	&0	&0.32346(4)	&0.0091\\
\hline
\hline
\end{tabular}
\begin{tabular}{|c|}
\hline
   Nb$_3$Pd$_{0.70}$Se$_7$  \\
\hline
\end{tabular}
\begin{tabular}{|c|c|c|c|c|c|c|}
\hline
  \hline
  Atom &  Site & Occupation & $x$ & $y$ & $z$ & $U_{\text{eq}}$\\
  \hline
Nb1	&4i	&1	&0.05848(5)	&1/2&	0.12765(3)&	0.0105\\
Nb2	&4i	&1	&0.11676(5)	&0	&0.26968(3)	&0.0083\\
Nb3	&4i	&1	&0.16833(5)	&1/2&	0.41178(3)&	0.0071\\
Pd1	&2a	&1	&0	&0	&0&	0.0126\\
Pd2	&2d	&0.393(5)	&0	&1/2	&1/2	&0.0109\\
Se1	&4i	&1	&0.35141(6)	&1/2&	0.49292(3)&	0.0089\\
Se2	&4i	&1	&0.16650(6)	&0	&0.06770(4)	&0.0108\\
Se3	&4i	&1	&0.40985(6)	&1/2&	0.09657(4)&	0.0103\\
Se4	&4i	&1	&0.22426(6)	&1/2&	0.20964(4)	&0.0091\\
Se5	&4i	&1	&0.28063(6)	&0	&0.35319(3)	&0.0078\\
Se6	&4i	&1	&0.47084(6)	&0	&0.23370(3)	&0.0083\\
Se7	&4i	&1	&0.02204(6)	&0&	0.37263(3)	&0.0088\\
\hline
\hline
\end{tabular}
\caption{Single-Crystal \emph{X}-ray diffraction collection and refinement: positional data for Nb$_2$Pd$_{0.67}$Se$_5$ and Nb$_3$Pd$_{0.70}$Se$_7$.}
\end{center}
\end{table}

{}

\end{document}